\documentclass{article}

\usepackage[numbers,sort&compress]{natbib}
\usepackage[final]{nips_2017_edited}

\usepackage[utf8]{inputenc} % allow utf-8 input
\usepackage[T1]{fontenc}    % use 8-bit T1 fonts
\usepackage{hyperref}       % hyperlinks
\usepackage{url}            % simple URL typesetting
\usepackage{booktabs}       % professional-quality tables
\usepackage{amsfonts}       % blackboard math symbols
\usepackage{nicefrac}       % compact symbols for 1/2, etc.
\usepackage{microtype}      % microtypography

\usepackage{graphicx}
\usepackage{amssymb} %maths
\usepackage{amsmath} %maths
\usepackage{calrsfs} 
\usepackage{amssymb} %for \emptyset

	\usepackage{boondox-cal}   %lowercase caligraphy \mathcal \mathbcal
\usepackage{bm} %for \bm boldmath of greek letters

\usepackage[labelformat = empty,position=top]{subcaption}%for top-left caption
\usepackage[export]{adjustbox}%for top-left caption%valign

\def\bfxhat{\widehat{\mathbf{x}}}

\def\bfx{\mathbf{x}}

\def\bfc{\mathbf{c}}
\def\bfe{\mathbf{e}}

\def\bfW{\mathbf{W}}

\def\bfWtilde{\mathbf{\widetilde{W}}}

\def\bfW{\mathbf{W}}
\def\half{\frac{1}{2}}
\def\bfphi{\bm{\phi}}
\def\bfpsi{\bm{\psi}}

\def\bfPsi{\bm{\Psi}}

\def\bfr{\mathbf{r}}
\def\bfrdot{\dot{\mathbf{r}}}
\def\bfs{\mathbf{s}}
\def\bfD{\mathbf{D}}
\def\bfF{\mathbf{F}}

\def\Wslow{\bfW^{\text{slow}}}
\def\Wfast{\bfW^{\text{fast}}}
\def\bfM{\mathbf{M}}
\def\bftheta{\bm{\theta}}

\def\bfu{\mathbf{u}}

\edef\ignore#1{}

\DeclareMathAlphabet{\pazocal}{OMS}{zplm}{m}{n}

\newcommand{\lcal}{\pazocal{L}}
\newcommand{\Mcal}{\pazocal{M}}

\newcommand*{\arialfont}{\fontfamily{phv}\selectfont}
\renewcommand{\O}{\emptyset}
\DeclareMathOperator{\Tr}{Tr}

\title{Learning arbitrary dynamics in efficient, balanced spiking networks using local plasticity rules}

\author{
  Aireza Alemi$^\dagger$\\
   Group for Neural Theory, ENS \\
   29 rue d'ulm, Paris--75005, France \\
  \texttt{$^\dagger$alireza.alemi@\{ens.fr, gmail.com\}} \\
  \AND
  Christian K. Machens\\
  Champalimaud centre for the unknown  \\
  Avenida Brasilia, 1400-038 Lisbon, Portugal \\
  \AND
  Sophie Den\`eve\thanks{Co-PI, alphabetic order} \\
  Group for Neural Theory, ENS\\
  29 rue d'ulm, Paris--75005, France \\
  \And
  Jean-Jacques Slotine$^*$\\
  Massachusetts Institute of Technology\\
   Cambridge, MA 02139, USA \\
}
\begin{document}

\maketitle

%TODO: Fig1B change \psi to \bfpsi
%TODO: Fig2 W -> W^\top
%TODO: specify \tau and the values use in the simulations

\begin{abstract}
	  Understanding how recurrent neural circuits can learn to implement dynamical systems 
	  is a fundamental challenge in neuroscience. The credit assignment problem, i.e. determining the 
	  local contribution of each synapse to the network's global output error, is a major obstacle in 
	  deriving biologically plausible local learning rules. Moreover, spiking recurrent networks 
	  implementing such tasks should not be hugely costly in terms of number of neurons and spikes, 
	  as they often are when adapted from rate models. Finally, these networks should be robust to 
	  noise and neural deaths in order to sustain these representations in the face of such naturally 
	  occurring perturbation. We approach this problem by fusing the theory of efficient, balanced 
	  spiking networks (EBN) with nonlinear adaptive control theory. Local learning rules are ensured 
	  by feeding back into the network its own error, resulting in a synaptic plasticity rule 
	  depending solely on presynaptic inputs and post-synaptic feedback. The spiking efficiency and 
	  robustness of the network are guaranteed by maintaining a tight excitatory/inhibitory 
	  balance, ensuring that each spike represents a local projection of the global output error 
	  and minimizes a loss function. The resulting networks can learn to implement complex dynamics 
	  with very small numbers of neurons and spikes, exhibit the same spike train variability as 
	  observed experimentally, and are extremely robust to noise and neuronal loss.

\end{abstract}

\section{Introduction}
%TODO: show the turning curves
%      any computation that can be formulated as dynamical system	
%      online local rule
%"A fundamental question in neuroscience is how upstream synapses (for example, the synapses between xi and hj in Fig. 1a) might be adjusted on the basis of downstream errors"
% TODO: cite DePasquale?, Burbank15, urbanczik and senn14

Recurrent networks in the nervous system perform a variety of tasks that could be formalized as dynamical systems. In many cases, these dynamical systems are learned based on examples (``desired'' trajectories), a form of supervised learning. For example, to learn to control an arm, sensory-motor circuits can learn to predict both the arm state trajectories and the sensory feedbacks, that are caused by specific motor commands.

Such learning occurs under several constraints. First, synapses have only access to local 
information. Because any local change in a synapse could have unpredictable effects on 
the rest of the network, previous approaches have often used non-local, biologically 
implausible learning rules such as temporal backpropagation \citep{werbos90} or 
FORCE learning \citep{sussillo09}.

Second, information in the nervous system is communicated with spikes. In order to obey 
the constraints imposed by the brain that is extremely costly to our metabolism \citep{laughlin98}, 
spiking neural networks should work with reasonably small number of spikes per dimension 
of the internal state dynamics. Such efficiency requires that there is no redundancy 
in the representation, so that spike trains of different neurons are uncorrelated.

Third, learning has to be able to resist various perturbations, such as varying levels 
of noise or the loss of neurons. Such robustness requires some level of degeneracy 
in neural populations, so that a given network has more neurons than strictly needed to learn a task.

No previous approach has been able to meet all three constraints at the same time. 
A few recent approaches based on adaptive control theory \citep{slotine86,slotine91} 
have been able to work with local learning rules, but 
%used very cost-inefficient spiking architectures
did not enforce spiking efficiency i.e., they do not require smallest possible number of spikes
\citep{dewolf16,gilra17}. Spiking efficiency and robustness 
were introduced in efficient balanced networks (EBN) \citep{boerlin13,deneve16}, 
but supervised learning in these networks has so far been limited to non-local rules \citep{memmesheimer14} or to linear dynamical systems \citep{bourdoukan15}.

In this study, we fuse the EBN framework \citep{boerlin13,deneve16} with adaptive 
nonlinear control theory \citep{slotine86,slotine91,sanner92} in order to derive 
local learning rules for arbitrary, nonlinear dynamics, while resulting in 
highly efficient and robust spiking networks. More specifically, we approximate nonlinear dynamics using a set of basis functions. The coefficients of the basis functions are then learned using a rule that utilizes correlation between the coefficients and the error signal (the difference between the desired trajectory and the approximated one). Adaptive control theory provides concepts such as Lyapunov functions and related theorems for quantifying converge properties of this rule. Based on these concepts we  
devise a local learning rule and construct an EBN that inherits these tools from adaptive control theory. 
\begin{figure}[!h]
	\centering       
    \begin{subfigure}[b]{0.5\textwidth}
	%\centering
	\Large{{\arialfont{A}}}
	%{\renewcommand{\familydefault}{phv}A}
	\includegraphics[width=\textwidth,valign=t]{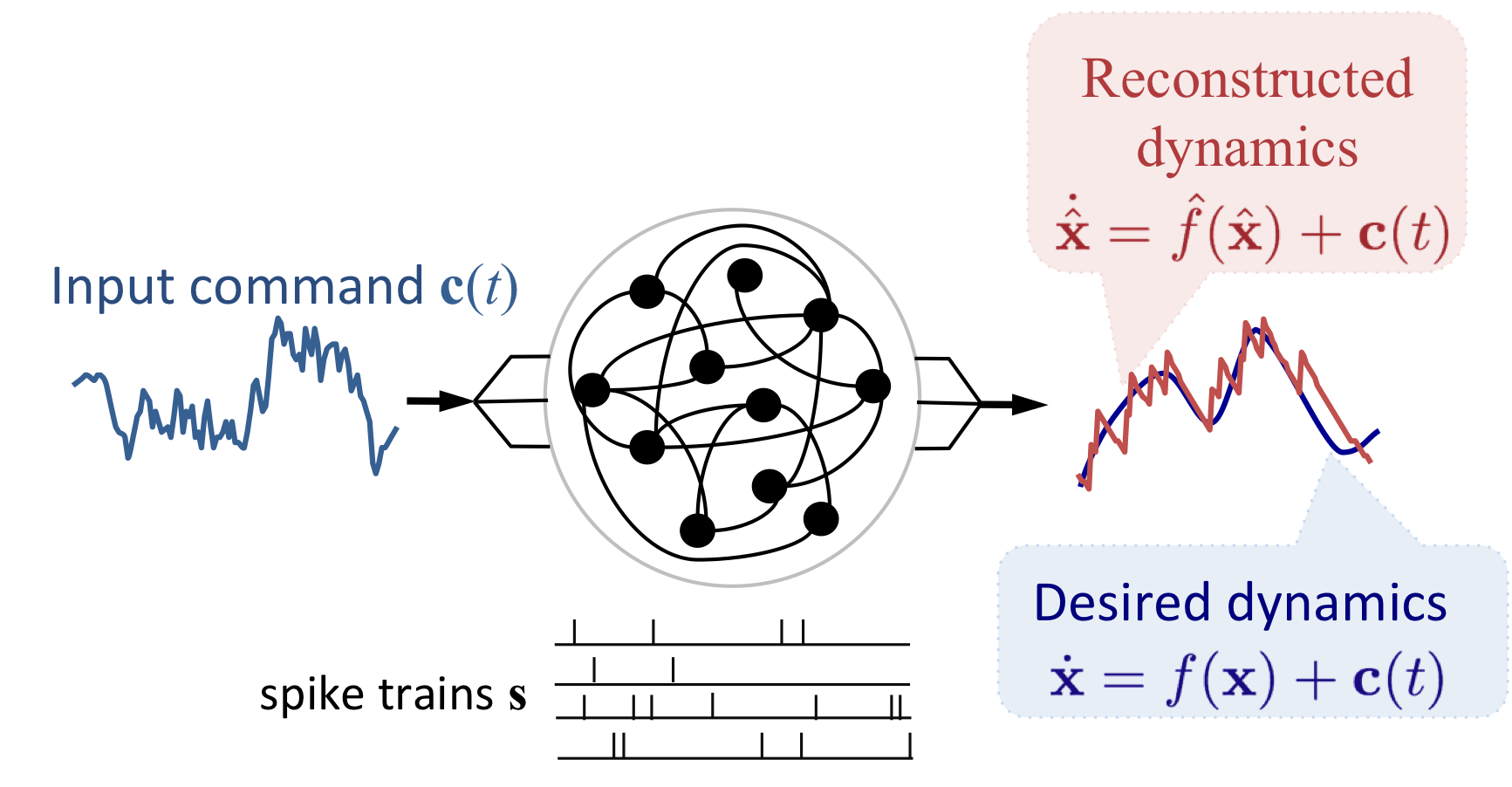}
\end{subfigure}
\hfill
\begin{subfigure}[b]{0.42\textwidth}
	\Large{{\arialfont{B}}}
	%\vspace{1em}{\large{B}}
	\includegraphics[width=\textwidth,valign=t]{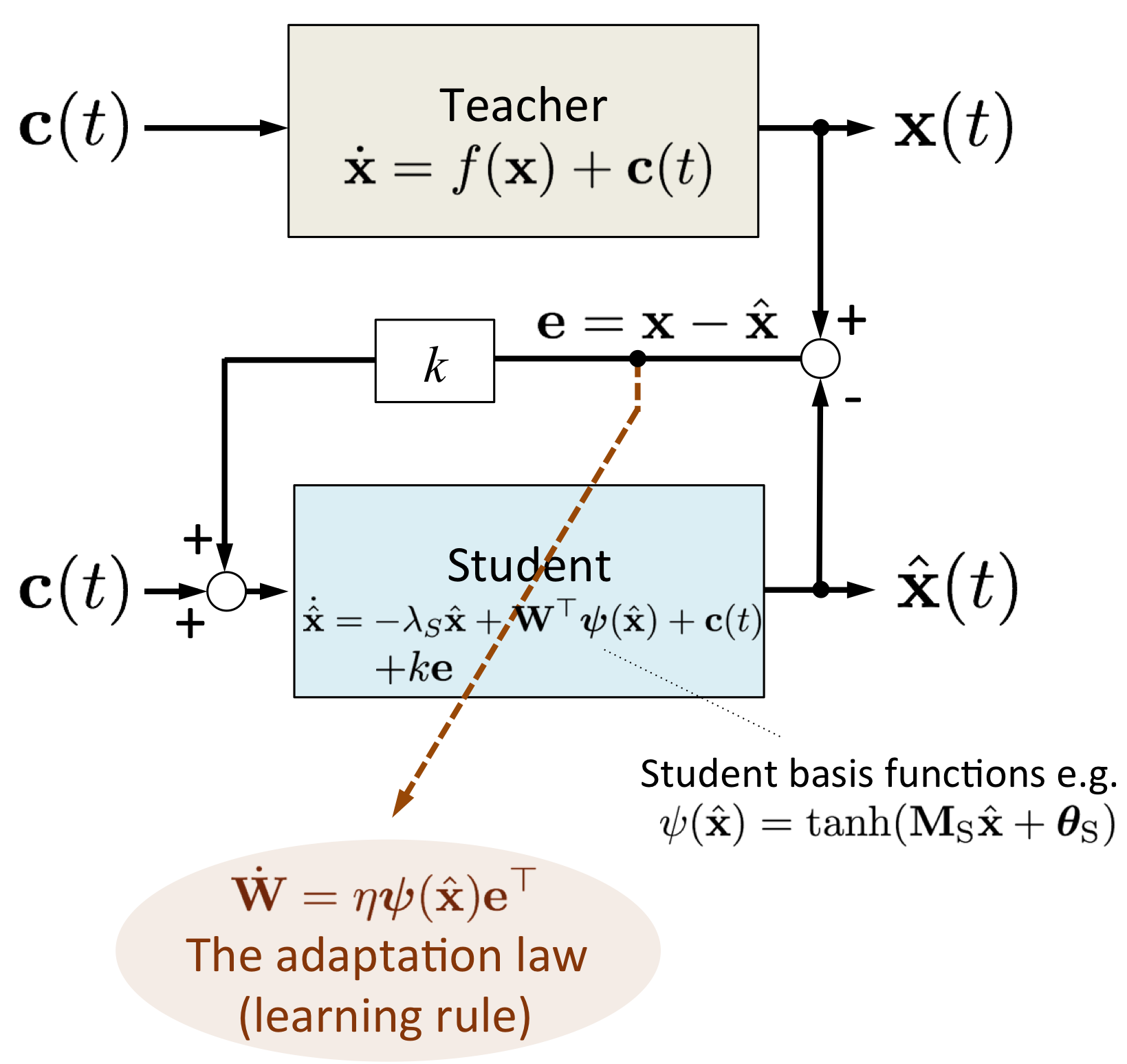}
\end{subfigure}

%	  	\internallinenumbers
	\caption{Schematics of the learning task and the control-theoretic solution. ~~A.~The task of learning an arbitrary 
		dynamical system. The network is presented with a random input command with certain statistics (cyan) and it 
		needs to produce a trajectory (red) close to the trajectory given by the teacher (blue). 
		The supervisory error signal, i.e. $\bfe=\bfx-\bfxhat$, should be 
		used to train the connections of the recurrent network to perform this task. ~~B.~Adaptive 
		nonlinear control theory for solving an estimation problem formulated as a teacher-student scenario.}
		\label{fig:problem}
\end{figure}

%\begin{figure}[!tbp]
%	\centering
%	%\begin{minipage}[b]{0.4\linewidth}
%	%		\centering
%	 %\begin{subfigure}[t]{0.45\textwidth}
%		%\large{A}
%		%\centering
%		\subfloat[]{\includegraphics[scale=0.08]{./problem.png}}
%		%\includegraphics[scale=0.11,valign=t]{./problem.png}
%	%\end{subfigure}
%	%\end{minipage}
%
%	\hfill
%
%	%\begin{minipage}[b]{0.4\linewidth}
%	%		\centering
%	%\begin{subfigure}[t]{0.22\textwidth}
%		%\centering
%		%\large{B}
%		\subfloat[]{\includegraphics[scale=0.08]{./controltheory.png}}
%		%\includegraphics[scale=0.1,valign=t]{./controltheory.png}
%	%\end{subfigure}
%	%\end{minipage}
%	%\hfill
%	\label{fig:problem}
%	\caption{The task of learning an arbitrary dynamical system. The network is presented by random input command (cyan) and it 
%		needs to produce the trajectory (red) close to the trajectory given by the teacher (blue). The error signal should be 
%		used to train the connection of the recurrent network to perform this task.}
%\end{figure}

\section{Adaptive nonlinear control theory: a teacher-student scenario}
The learning task that we consider is presented in Fig.~\ref{fig:problem}A.
An input signal $\mathbf{c}(t)$ drives a recurrent spiking network. The task is to learn the network 
connectivity such that a desired dynamics (in blue with the state variable $\bfx$) 
can be decoded from the network spike trains (in red with the reconstructed state $\bfxhat$). 
This task could correspond to a sensorimotor learning task, where the network learns a 
\emph{forward internal model} that receives the \emph{efference copy} 
(i.e. a copy of the input motor command) to predict the body position given the sensory error feedback 
and the motor command \citep{wolpert00}.

More formally, we assume that the desired dynamics stem from a ``teacher'' dynamical system of the general form 
\footnote{Our dynamic variables $\bfx(t)$ and input 
	signals $\bfc(t)$ are a function of time but sometimes for simplicity we omit the time variable $t$, writing 
	them as $\bfx$ and $\bfc$} %=\frac{d\mathbf{x}}{dt}
\begin{equation}
% \dot{\bfx} = - \bfx + f(\bfx) + \bfc(t) 
 \dot{\bfx} = f(\bfx) + \bfc(t) 
 \label{eq:generalform}
\end{equation}
in which $\bfx$ is a time-dependent dynamic vector of continuous, real-valued variables,  $\bfc(t)$ is a 
time-varying input or command signal chosen as a filtered random signal unless otherwise stated, 
and $f(.)$ is an arbitrary, vector-valued function.
%and $\bfM_T$ and $\bftheta_T$ are parameters of the basis functions chosen randomly from a Gaussian distribution.
Note that if the teacher dynamical system is of higher order, we can transform it into a 
higher-dimensional, first-order dynamical system. 

To understand our approach to learning this teacher system, we will first ignore the neural network, and briefly recapitulate how the problem is solved in 
adaptive nonlinear control theory \citep{slotine86,slotine91,sanner92}. Here, the goal is to estimate the parameters of a ``student'' dynamical system such that its dynamics matches the dynamics of the ``teacher'' system (See Fig.~\ref{fig:problem}B). The student system has the following form:
\begin{equation}
	\dot{\bfxhat} = -\lambda \bfxhat +  \mathbf{W}^{\top} \bfpsi(\bfxhat) + \mathbf{c}(t) + k \, \mathbf{e},
	\label{eq:student}
\end{equation}
 where  $\bfxhat$ is the dynamic vector variable of the student, $-\lambda\bfxhat$ is a leak-term, $\bfW$ are the parameters of the student 
 dynamics, $\bfpsi(\bfxhat)$ are the student basis functions (e.g. $\bfpsi(\bfxhat)=\tanh(\bfM_\text{S}^\top\bfxhat + \bftheta_\text{S})$),
 $\bfe=\bfx-\bfxhat$ is the tracking error, and $k$ is the feedback gain. The goal of learning is to adapt the parameters $\bfW$ such that the tracking error is minimized. In the initial phases of learning, the feedback gain $k$ is usually set to a large value, which causes the student to follow the desired teaching dynamics closely. In turn, the student variable $\bfxhat$ already traverses the state space correctly which provides the opportunity for adjusting the adaptive parameters $\bfW$ in order to learn the dynamics. Once these parameters have been learned (at least approximately), the feedback gain is decreased, which also helps to get good generalization behavior and reduce the number of training iterations. Once the system is learned, the feedback gain can be set to zero, which is what we did in the test phases ($k_\text{test}=0$). 

If the student has sufficiently many rich basis functions to approximate the function $f(.)$, then there exists a solution called $\bfW^\text{true}$. 
If we define a Lyapunov function as 
   \begin{equation}
	   V  =  \half \bfe^{\top} \bfe  +  \frac{1}{2\eta} \Tr\big(\bfWtilde^{\top} \bfWtilde\big),
   \end{equation}      
   where $\bfWtilde=\bfW^\text{true}-\bfW$ is the estimation error and $\Tr(.)$ is the matrix trace operator, 
   it can be shown (see Supplementary Materials) that the 
   following adaptation law (or the learning rule) 
   \begin{equation}
	   \dot{\bfW} = \eta \, \bfpsi(\bfxhat)  \mathbf{e}^{\top},
	   \label{eq:adaptlaw}
   \end{equation}
    where $\eta$ is a learning rate, will decrease the Lyapunov function, i.e.   $\dot{V} = - (k+1)\, \bfe^{\top} \bfe \le 0$. 
    This result together with boundedness conditions for $V$ and  $\ddot{V}$ guarantees that $\dot{V}$, 
    and hence the tracking error $\bfe$, will asymptotically go to zero and that the system is 
    asymptotically stable \citep{slotine91}. Moreover, if the input $\bfc(t)$ is rich enough, 
    then $\bfW\rightarrow\bfW^\text{true}$. 
    
%It should be noted that though we considered first-order form in Eq.~\ref{eq:generalform} 
%adaptive control theory can still work by making the error feedback richer: 
% if the teacher dynamical systems is given as an $n$-order system, a fundamental result of adaptive control
%theory \citep{slotine86} states that a linear combination of the error and its derivatives
%up to $\bfe^{(n-1)}$ i.e. $\Scal = \sum_{i=0}^{n-1}\lambda_i \bfe^{(i)}$ can be used instead of
%just $\bfe$ to drive the student and to estimate its parameters. We are now equipped with these tools in order to
%come up with a learning rule for a spiking network.

%\begin{figure}
%	\centering
%	\includegraphics[scale=0.1]{controltheory.png}
%	\label{fig:controltheory}
%	\caption{Adaptive nonlinear control theory for solving an estimation problem formulated as a teacher-student scenario}
%\end{figure}

\section{Learning a functional spiking network}
\begin{figure}[!h]
	\includegraphics[scale=0.12]{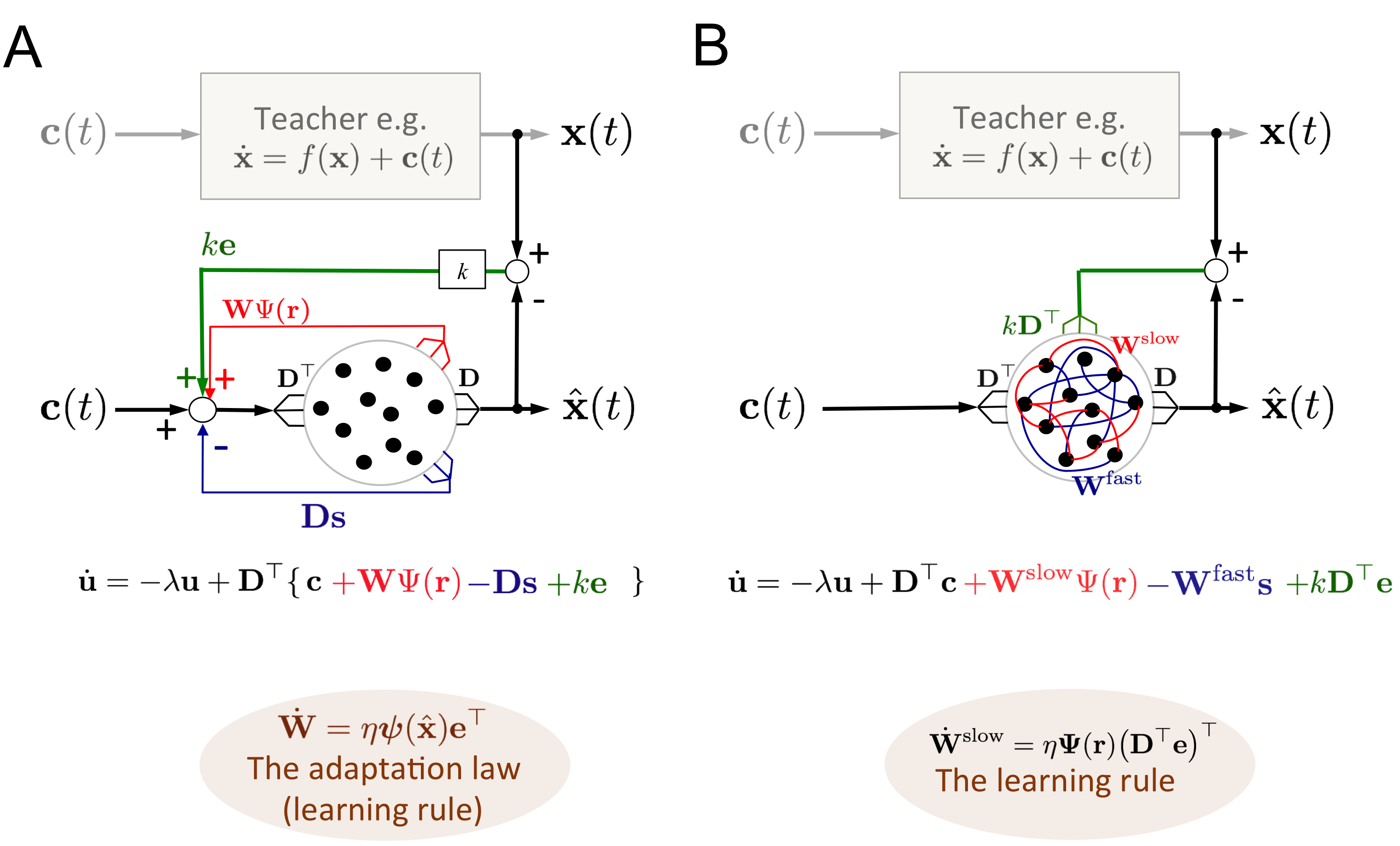}
		  	%\internallinenumbers
	\caption{Building a spiking network approximating arbitrary dynamical systems. 
		~~A.~The unfolded version of the network of LIF neurons 
		(with membrane potentials $\bfu$ and a threshold and reset mechanism) with 
		three feedback loops implementing desired dynamics 
		of $\bfx$ in the network: the role of red loop is to implement nonlinear dynamical system, 
		the blue one provides efficiency 
		and robustness, and the green one feeds the error back into the network such that $\bfxhat$ 
		closely tracks $\bfx$. 
		~~B.~The folded version of the network with slow and fast connections. 
		The corresponding learning rules are shown 
		at the bottom.}
		\label{fig:network}
\end{figure}

We want to translate our student into a recurrent network of $N$ leaky integrate-and-fire (LIF) neurons. Previous work has shown how to implement arbitrary linear or nonlinear dynamical systems in efficient balanced networks \citep{boerlin13,thalmeier16}. EBN theory is based on two assumptions. First, an estimate of the $K$-dimensional state variable, $\bfx$, can be extracted from the filtered spike trains, $\bfr$, using a linear decoder, such that
\begin{equation}
	\bfxhat=\bfD\bfr,
	\label{eq:deocoder}
\end{equation}
where $\bfD$ is a fixed decoding weight matrix of size $K\times N$ and the relation between $\bfr$ and the spike trains $\bfs$ is $\bfrdot = -\lambda\bfr + \bfs$.  Second, neurons in the network fire spikes such that this estimate, $\bfxhat$, closely follows the true state variable, $\bfx$, under cost constraints. Specifically, the network minimizes the following objective function, 

\begin{equation}
	\lcal = \langle L \rangle
	=
	\Big\langle \|\bfx - \bfxhat\|^2 + \mu{\|\bfr\|}_2^2+\nu\|\bfr\|_1\Big\rangle.
	%= \langle\lcal(t)\rangle
	\label{eq:loss}
\end{equation}

where $\Vert\cdot\Vert_p$ denotes the L$_p$-norm, $\langle\cdot\rangle$ is an average over time, and $\mu$ and $\nu$ determine the costs associated with spiking. The first term on the right-hand side of Eq.~\ref{eq:loss} ensures a good reconstruction, whereas the L2 and L1 cost functions on neural activity ensure a distributed and sparse spiking activity, respectively \citep{boerlin13}.

The resulting network consists of LIF neurons whose membrane potentials, $u_i$'s, obey the equation \citep{boerlin13,bourdoukan12} (see also Supplementary Materials)

\begin{equation}
	%\dot{\bfu}  =  -\bfu + \bfD^\top (\bfx + \dot{\bfx}) - \Wfast\bfs, \label{eq:vdyn}
	u_i = \bfD_i^\top  (\bfx - \bfxhat ) - \mu r_i,	\label{eq:veq}
\end{equation}

and whose firing thresholds are given by $T_i = \frac{1}{2} (\|\bfD_i\|^2_2+\mu + \nu)$, where $\bfD_i$ is the $i$-th column of $\bfD$. The thresholds ensure that each neuron fires only when its spike decreases the objective function  (Eq.~\ref{eq:loss}). %, and the ``fast'' recurrent connections, $\Wfast = \bfD^\top\bfD + \mu\mathbf{I}$, where $\mathbf{I}$ is the identity matrix, make sure that the membrane potentials properly track the objective function. Indeed, the voltage of each neuron is simply a projection of the reconstruction error so that 
%$\bfu = \bfD^\top  (\bfx - \bfxhat ) - \mu \bfr$. 
 Differentiating Eq.~\ref{eq:veq} and simplifying it yields the following membrane potential dynamics
\begin{equation}
\dot{\bfu}  =  -\lambda\bfu + \bfD^\top\bfc -  (\bfD^\top \bfD +\mu \mathbf{I})\bfs + \bfD^\top ( \lambda\bfx + f (\bfx) ),
\label{eq:vdyn}
\end{equation}
which shows the optimal weight matrix for the fast connections is $\Wfast = \bfD^\top \bfD +\mu \mathbf{I}$, where $\mathbf{I}$ is the identity matrix, and the optimal encoding weight matrix is $\bfD^\top$.
%In order to implement arbitrary dynamical systems in this framework, given by $\dot{\bfx} = f(\bfx) + \bfc$, we simply replace the term $\dot{\bfx}+\bfx$ on the right-hand-side of Eq.~\ref{eq:vdyn}, to obtain
%\begin{equation}
%\dot{\bfu}  =  -\bfu + \bfD^\top\bfc -  \Wfast\bfs + \bfD^\top f (\bfx).
%\end{equation}
In order to implement arbitrary dynamical systems in this framework, we need to include two approximations. First, we can approximate $ \lambda\bfx + f (\bfx)$ as a sum of the ``student'' basis functions (similar to Eq.~\ref{eq:student}), so that $ \lambda\bfx + f (\bfx) = \sum_i \bfW_i \phi_i (\bfx)$, where $\bfW_i$'s are the columns of $\bfW$, and $\phi(.)$ can be any sigmoidal nonlinearity set to $\tanh(.)$ in the simulations.     Second, we can self-consistently replace the state $\bfx$ by its estimate based on network activity (Eq.~\ref{eq:deocoder}). In turn, we obtain (Fig.~\ref{fig:network}B)

\begin{equation}
	\dot{\bfu} = -\lambda\bfu + \bfD^\top \bfc- \Wfast\bfs + \bfD^\top \bfW^\top \bfphi(\bfD {\bf r}).
	\label{eq:studentNet}
\end{equation}

From a biological point-of-view, the last term corresponds to 
nonlinear and highly structured dendrites. There is evidence that dendrites can perform nonlinear operations \cite{poirazi03arith,poirazi03pyr}. While dendrites are often nonlinear, the required detailed spatial organization of their inputs
is rather speculative. To relax this latter constraint, we can take advantage of the fact that the network's internal state dynamics is extremely robust to large amounts of noise (as ensured by its constant, greedy minimization of its own coding errors), as long as $N > 2K$. Thus, we can replace $\bfphi(\bfD {\bf r})$ with an unstructured, nonlinear dendrite that receives random connections from other neurons,
$\Psi_i(\bfr) = \bfphi (\bfM_i^\top \bfr + \bftheta_i)$ where $\bfM_i$'s are the columns 
of the matrix $\bfM$. Indeed, the quantity $\bfM^\top\bfr$ can be written as the sum of 
its projection onto the decoder $\bfD$, and its projection onto the null space of $\bfD$ (denoted by $\bfD^\O$), so that ${\bfM^\top \bfr = \Mcal^\top \bfxhat + \Mcal^\top \bfD^\O \bfr}$. Given that EBNs do not control spiking in the null space of the decoder, the second term acts as unstructured noise. \footnote{We note that it would be possible to train the parameters $\bfM$ using unsupervised learning, since $\bfD^\top$ corresponds to the eigenvectors with non-zero eigenvalues 
in the neural correlation matrix. This should render the network even more efficient and robust as will be explored in the future.}

To clarify the learning problem, we will write the resulting student network dynamics in the general form
\begin{equation}
	\dot{\bfu} = -\lambda\bfu + \bfF \bfc- \Wfast\bfs +\Wslow\bfPsi(\bfr).
	%\label{eq:studentNet}
\end{equation}
We note that the network has three sets of synaptic connections. The feedforward connections, $\bfF$, receive and weight the external signal input, $\bfc$. 
The fast connections, $\Wfast$, guarantee the proper and efficient distribution of spikes across the network. Finally, the slow connections, $\Wslow$, implement the dynamics of the student system. In previous work, it is shown how to learn the feedforward connections \citep{brendel17} and the fast recurrent connections \citep{brendel17,bourdoukan12}. Though all connections could be trained simultaneously, we here concentrate on how to train the slow connections based on example trajectories from an unknown (teacher) dynamical system. We use a fixed random decoding weight matrix $\bfD$ (which is close to optimal in the case of uncorrelated command signals) and set the feedforward connections and recurrent connections to their optimal values, $\bfF=\bfD^\top$ and $\Wfast = \bfD^\top \bf{D} +\mu \mathbf{I}$.

%When the membrane potential of neuron $i$, denoted by $\bf{u_i}$,
%is a leaky integrator $\dot{\bfu}=-\lambda\bfu + $,
%reaches a threshold $T_i$, the neurons spike and then reset to a reset potential $u_\text{reset}$.

A direct translation of the adaptive control to the neural network will finally permit us to 
define a local learning rule for the slow connections. Rather than feeding back the error into the student network by 
adding it to its input, we directly inject the errors as feedback to each neuron with connections $\bfD^\top$, which is 
mathematically equivalent. In the presence of this feedback control loop, the network equation becomes
\begin{equation}
	\dot{\bfu} = -\lambda\bfu + \bfF \bfc- \Wfast\bfs +\Wslow\bfPsi(\bfr) + k\bfD^\top\bfe.
	\label{eq:studentNetWe}
\end{equation}

Moreover $\Wslow = \bfD^\top\bfW^\top$ is directly related to the coefficients $\bfW$ 
of the basis functions in the control theory framework. This allows us to map the adaptation law 
(Eq.~\ref{eq:adaptlaw}) for the coefficients of the basis function to the following learning rule 
for the slow connections. Replacing $\bfe$ in
Eq.~\ref{eq:adaptlaw} with $\bfD^\top\bfe$, we obtain
\begin{equation}
	\dot{\bfW}^\text{slow} = \eta \, \bfPsi(\bfr)  (\bfD^{\top}\mathbf{e})^\top.
	\label{eq:learningrule}
\end{equation}
In other words, the resulting learning rule is the product of  a pre-synaptic input (passed through the dendritic nonlinearity) 
and the projection of error feedback. Hence, the learning is local.

%In our implementation, we used $\bfPsi(\bfr)=\tanh(\bfM\bfr+\bftheta)$ where $\bftheta$ is sampled from a Gaussian 
%distribution and $\bfM$ is either random or has a lower rank as $\bfM=\bfM'\bfD\bfr$ 
%in which $\bfM'$ is chosen randomly. Note that
% the parameters of the basis functions $\bfPsi(.)$ are not adaptive and fixed during the learning process.
% We also incorporated a constant and linear basis functions to make the function approximation easier. 

For the case of sensorimotor learning, the quantity $k\bfe$ could correspond to visual or somatosensory ``prediction errors'', 
e.g. the difference between the sensory input and its prediction based on the efference copy of the motor commands. 
Over the course of learning, the errors become smaller and would eventually vanish in the absence of motor 
noise or sensory noise. Consequently, the feedback would become silent and could be removed entirely.  

The general idea of this derivation, and the mapping from the control-theoretical framework onto efficient balanced networks, is also illustrated in Fig.~\ref{fig:network}A.

\begin{figure}[!h]
	\centering
		\includegraphics[scale=0.12]{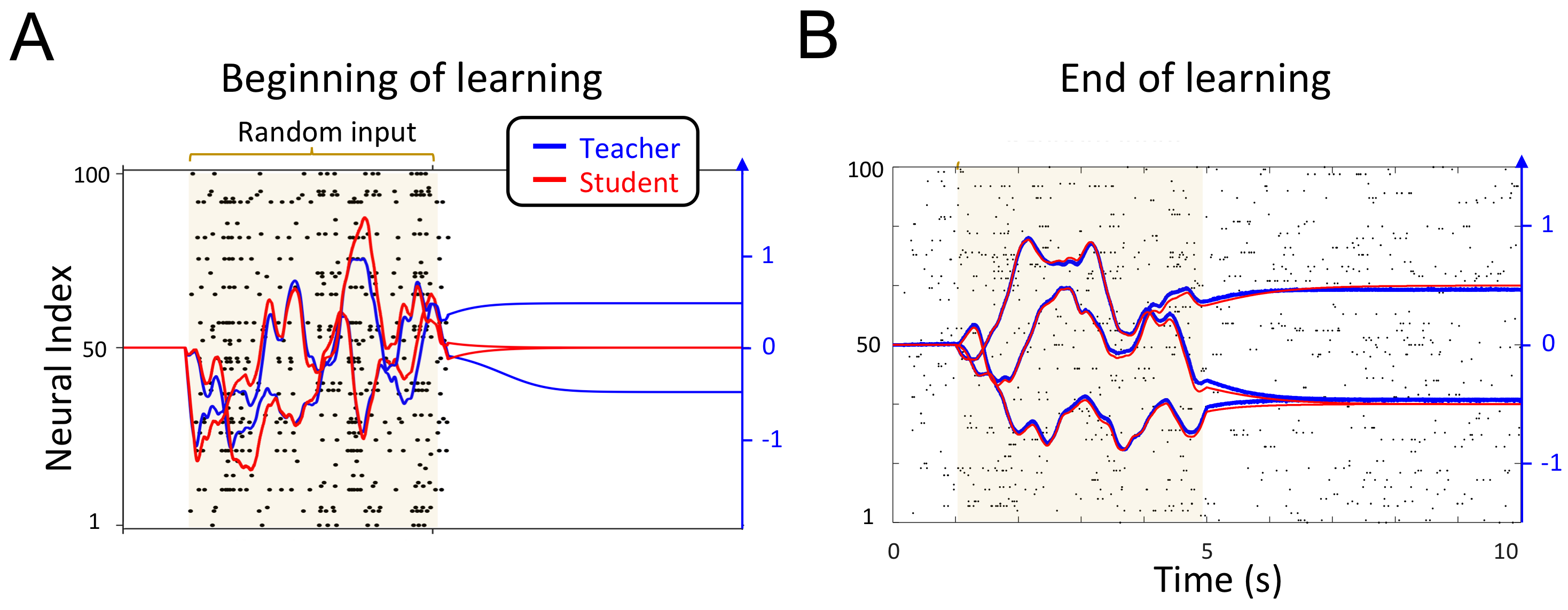}

	%\internallinenumbers
	\caption{Result of learning a bistable attractor in an spiking network. 
		The desired dynamics given by a teacher 
		is $\dot{x}=x(0.5-x)(0.5+x) + c(t)$ where $c(t)$ is a random input signal.~~A.~In the beginning of learning, 
		the reconstructed dynamics of the student network (in red) during the presentation of the input (shaded area) 
		is due to the random inputs (two trials shown). After the input is turned off, the reconstruction does not go to any of 
		the attractor states.~~B.~At the end of learning, when the input is off, the reconstructed 
		signal falls into one of the stable attractors (three trials).}
		\label{fig:bistable}
\end{figure}

\begin{figure}[!h]
	\centering
	\includegraphics[scale=0.8]{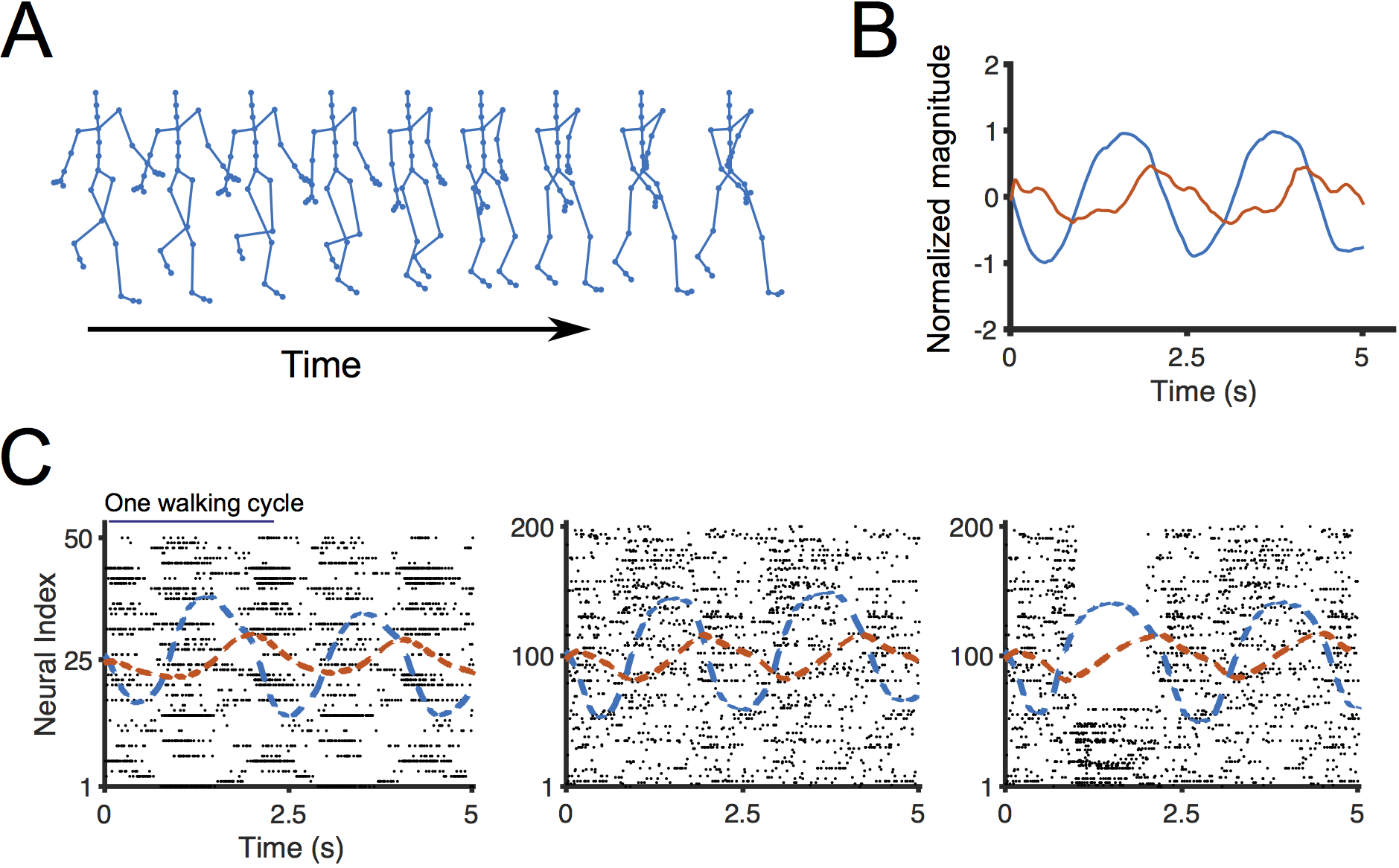}
	
	%\internallinenumbers
	\caption{Result of learning to walk in spiking networks.~~A.~The dynamics of the walk of a sticky man implemented by a 
		network of spiking model neurons. The training data is from CMU Motion Caption Library.~~B.~Two first 
		principal components of the desired trajectory of the walk. The original dataset for the 
		walk has 62 channels.  ~~C.~Raster plots for networks that learned a walking dynamics based on 
		four principal components of a trajectory of the walk. The left panel shows the plot 
		for a network of 50 neurons. Four principal components were provided to the network as desired trajectories. The reconstruction of the first two principal components are shown overlaid on the raster plot. As the number of neurons in the network increases to 200 neurons 
		(middle panel) the firing rates decrease and spiking activity becomes more asynchronous and 
		irregular and the reconstruction improves. The network is drastically robust to silencing neurons as shown in the right-most panel 
		where $70\%$ of the neurons are silenced for a period of around one second but the network continues to generate walking with a short, negligible distortion. 
		Note that after silencing the neurons, the network weights do not undergo any learning, 
		but other active neurons compensate for neuronal loss in the network.}
	\label{fig:walking}
\end{figure}

\section{Implementing nonlinear dynamics}
We employed the framework described above in the following example tasks.

\subsection{Bistable attractor}
The bistable attractor is an important nonlinear dynamical system which is widely studied in systems neuroscience.  
We implemented the following one-dimensional attractor dynamics
\begin{equation}
	\dot{x} = x(0.5-x)(0.5+x) + c(t),
\end{equation}
where $c(t)$ is a random input command.
This system has two stable fixed-point attractors at $x=\pm0.5$ and a saddle-point at $x=0$. We used $N=50$ 
neurons with random readout weights $\bfD$ to implement this system. As shown in Fig.~\ref{fig:bistable}A,  the random input $c(t)$ 
drives the student network in the beginning of learning. Once this input is zero the desired dynamics given by the teacher (in blue)
goes to one of the stable attractors. But the decoded dynamical variable of the network (in red) has not yet 
learned the desired attractors. After learning (Fig~\ref{fig:bistable}B), the network closely follows the desired 
trajectories in response to a random unseen input (not used during training) and, more importantly, 
once the input is turned off, it settle on one of the attractors. Note that the network exhibits 
realistic spiking activity with low firing rates and asynchronous, irregular spiking activity. 
The overall number of spikes is drastically lower than that of spiking 
networks that effectively use \emph{rate codes} \citep{eliasmith04,eliasmith05,gilra17}.

\subsection{Motion capture example}
We also implemented the walking dynamics from the database of Carnegie Mellon University 
Motion Capture Library (MOCAP) \url{http://mocap.cs.cmu.edu/}. We used a similar preprocessing procedure as in \citep{sussillo09}. Briefly, 
the data was taken from file ``08\_01.amc'' which consisted of 62 channels out of which three were set to zero, 
converting the movement to a walk on a treadmill. We preprocessed the data by a moving average.
 
In order to reduce the dimensions of the input signal, we took the first four
 principle components. This reduced
input closely follows the original walking dynamics with minor loss. 
As the system is a mechanical system, and therefore of second order,
we require both velocity and position information. Therefore, in order to be able to model it, we fed both the
trajectories and its derivatives to the network, so that the overall input was 8-dimensional. 
Similarly, we provided a linear combination of the position error and the velocity error as the error feedback to the network.
We then learned the dynamics of the 8-dimensional input with two different networks, one of size $N=50$ and and one of size $N=200$ (compare with \citep{sussillo09}). 

Fig.~\ref{fig:walking}A shows the learned walking, and the left and middle panels of Fig.~\ref{fig:walking}B shows 
the raster plot of the 50-neuron and 200-neuron networks, respectively. An important feature of our model is its 
expansion: the number of input channels ($K$) must be smaller than 
the number of neurons ($N$) in the network. In order for the network to have the desired aforementioned
properties, as a rule of thumb, the expansion ratio ($\Lambda=N/K$) should be larger than 5--10. This expansion provides many 
possible solutions for the network to perform a task. Thanks to the efficiency 
principle, the network is able to choose the most efficient one. It should be noted that the realistic 
spiking activity is closely linked to the expansion: increasing the expansion ratio results in lower 
firing rate and more irregular spiking activity (compare left and right panels of Fig.~\ref{fig:walking}C).
This is one of the main differences with other spiking networks (but in fact they use a \emph{rate code}) 
implementing complex dynamics such as the Neural Engineering Framework (NEF) \citep{eliasmith04,eliasmith05}, as they would
typically need firing rates in the order of the inverse of synaptic time-scales to learn properly.
Another striking feature of the model is its robustness to neural death and noise. Thanks to the presence of
the fast connections, even if $\%70$ of the neurons in the network are silenced the network would still be 
able to perform the task with minimal loss in the performance (right-most panel in Fig.~\ref{fig:walking}C). 
This is consistent with the expected robustness of the EBN framework \citep{barrett15}.

%  \annote[AA]{If the teacher dynamical systems is given as an}{where to include this paragraph? 
%  	How to justfy using edot for walking?} $n$-order system, a fundamental result of adaptive control 
% theory \citep{slotine86} states that a linear combination of the error and its derivatives 
% up to $\bfe^{(n-1)}$ i.e. $\Scal = \sum_{i=0}^{n-1}\lambda_i \bfe^{(i)}$ can be used instead of 
% just $\bfe$ to drive the student and to estimate its parameters. We are now equipped with these tools in order to 
% come up with a learning rule for a spiking network.

\section{Discussion}
  %- Nice properties...
   We have proposed a local learning  rule in an spiking neural network of LIF neurons for learning 
   arbitrary complex dynamics. The resulting
   networks exhibit low firing rates with asynchronous, irregular spiking activity 
   where the spiking representation is as efficient as
   possible. The efficiency principle that we have exploited has direct consequences: the inhibitory input currents in
   each neuron closely track the excitatory input (E/I balance); and the network is highly robust to noise,
   neural elimination, and uncertainty in desired dynamics \citep{boerlin13,barrett15}.
   The learning rule is obtained thanks to concepts and
   tools in nonlinear adaptive control theory. This approach has the benefit of providing a systematic way
   to study convergence and stability properties of the learning process which currently is not pursued in the mainstream learning 
   procedures in neuroscience for spiking neurons \citep{sussillo09,abbott16,thalmeier16}. 
   %The neuron model can become more realistic as Hodgkin-Huxley model has been used for implementing an EBN network \citep{schwemmer15}. 
   Studying the effect of synaptic 
   delays needs to be addressed in future versions of our model, although previous work suggests that increasing the amount of noise 
   may help avoiding oscillation and synchronization \citep{chalk16}.
   
   The learning rule can a learn any complex dynamics $f(.)$. However, it may not learn the target dynamics if the input does not have “sufficient richness” condition (see \citep{slotine91}). If function $f(.)$ has singularities that make the dynamics unstable, the learning may fail. Another requirement is that the states (for example, position and velocity in the motion capture example) need to be well-defined. Any task that can be cast as a well-behaved dynamical system can be learned in our framework.
   
The different parts of the spiking network have straightforward biological interpretations.
Fast connections could correspond to interneurons which would mostly 
be driven by monosynaptic, fast AMPA synapses from excitatory neurons, 
targeting the soma and relying on ionic (GABA-A) neurotransmission. 
Slow connections could be implemented by slower metabolic channels 
(e.g. NMDA/GABA-B) and correspond either to 
direct connections between pyramidal cells, or disynaptic inhibition using another 
type of interneurons, in the case of negative weights. 
   
   The network seems to be highly structured in our framework. However,
    it has already been shown that most of 
   the other connections in the network can be trained using local 
   spike-time-dependent plasticity rules with the exception of $\Wslow$ which is the contribution of the present work. 
   In particular, the fast connections $\Wfast$ can be trained using local plasticity 
   rules, while the feedforward connections (and thus, the decoding weights) can be trained using a 
   Hebbian spike-time-dependent rule \citep{brendel17,bourdoukan12} . Note that these inhibitory fast connections 
   could be thought of as implementing a kind of `dynamic' Winner-Take-All (WTA) network where once a
   neuron wins and spikes, then other neurons have a chance to fire as well. This is
   a different scheme from a distributed WTA of inhibitory neurons \citep{wangslotine06}.
 
%   One of the strong predictions is that neural circuits must be robust to noise and neural elimination \citep{barrett15}. This could
% be tested by inactivation of neurons using experimental methods such as optogenetics. We predict that inactivating a portion of a
% neural population during the performance of a complex movement task should not strongly degrade the performance, 
% but that spared neurons will compensate by changing their tuning curves to compensate (which fits with recent results in \citep{liSvobodaDruckmann16}).
   
 In our general framework, nonlinear dendrites compute the basis functions where we used $\tanh(.)$ nonlinearity. However, the learning rule is not limited to that and can work with any basis functions (nonlinearities) that can approximate arbitrary functions e.g. Gaussian, sigmoid basis functions, though sigmoids are more biologically plausible. Apart from the form of nonlinearity, we have a large degree of freedom in the choice of parameters of basis functions $\bfM$ i.e., in the design of the dendritic tree. The case that might be easiest to map to the control-theoretic framework is the case where $\bfM=\bfD\bfM'$ 
is a low rank matrix. But other cases such as random $\bfM$ also work in practice. We provided intuitions 
for why such a suboptimal architecture still functions: the network activity is forced to be as efficient 
as possible due to the presence of the fast connections (E/I balance). This defines an optimal 
combination of neural activities for each state $\bfxhat$. In other words, neural firing rates 
become almost completely determined by the internal state estimate (up to their ``poisson-like'' variability). 
This neural activity projected onto an arbitrary matrix will also be a function of $\bf \hat x$, 
plus some unstructured input that may be considered as noise. This noise is automatically compensated 
by the network robustness. In the case of the walking example, we tested a diagonal matrix  $\bfM$ 
(in which case, the dendritic nonlinearity is replaced by a nonlinear transfer function for each neuron) 
and the network was still able to learn without any difficulty.  The best choice of 
$\bfM$ (including the possibility of learning these parameters) remains to be explored further elsewhere. 
Our approach is aligned with other works taking similar strategies for implementing 
nonlinearities in recurrent networks \citep{abbott16,thalmeier16}.
 
  % expansion, asynchronous irregular firing,...
% Using nonlinear adaptive
%   control theory, one can come up with a local learning rule for the Neural Engineering Framework to learn complex dynamics \citep{gilra17},
%   however these networks don't have the fast connections to provide spiking efficiency and balance. More importantly, 
%   they do not scale up easily as they would need overall lots of spikes to implement dynamics 
%   \citep{sussillo09,gilra17}. They are not 
%   able to exploit the expansion completely, therefore it would be more challenging 
%   for these networks to exhibit realistic spiking activity and robustness. 
%   Our work is different, as it seamlessly blends and exploits earlier work on 
%   EBNs, which exploited the spiking nature of neural activity (rather than treating it as a hindrance).
  	
  Nonlinear adaptive control theory yields  a local learning rule for the Neural
  Engineering Framework \citep{gilra17} to learn complex dynamics. However, the resulting networks do not implement
  fast connections to provide spiking efficiency and balance. This in turn implies that they do not
  scale easily when applied to large dynamics \citep{sussillo09,gilra17}, making it more challenging for these networks to
  exhibit realistic spiking activity and robustness. Our work aims to seamlessly blend nonlinear adaptive control with
  earlier work on EBNs, which exploits  the spiking nature of neural activity rather than treating it as a hindrance.	
  	
 %- difference with other control theoretic approaches: optimal control, NEF, 

  %- Neuroscience preditions: robustness and deactivation of neurons in experiments, nonlinear dendrites, imlementation on more realistic neurons...
  %TODO: lim13?
   In conclusion, we argue for a close relationship between spiking efficiency, robustness and 
   the tight E-I balance observed in cortical circuits.  We suggest that experimentally observed 
   spike trains, with low-firing rate and asynchronous, irregular spike trains, are a signature 
   of an efficient spike-based coding, and not a noisy rate-based population code. The network 
   effectively implements dimensionality reduction: regardless of its size, the dimensionality 
   of its population dynamics and recurrent weights eventually becomes restricted to the 
   dimensionality of the task, while neural fluctuations occur in direction orthogonal 
   to the task. Thus, we predict that such low-dimensional dynamics emerge through 
   experience in biological neural circuits. Finally, learning in biological circuits 
   would require that feedback connections monitoring the network performance both drive 
   the neurons and modulate learning. Each slow connection is learned as a function of 
   the correlation between presynaptic input rate and postsynaptic error feedback, 
   until this error feedback is canceled. Thus, only neurons with correlations to 
   the error (and presumably contributing to such error) see their synaptic weights change. 
   In contrast, backpropagation would result in diffuse change in the entire network. 
   These predictions have broad implications for Brain Machine Interfaces. In the near 
   future, this theory may pave the way for implementing more complex tasks and for 
   spike based unsupervised, hierarchical and reinforcement learning, in both biological 
   and artificial spiking networks.

  %- Toward universal computation
  %\citep{shannon41,bournez16}
    
  %- Application: light-weight robots (DARPA), control instead of estimation
  The framework presented here can have engineering applications for example in light-weight robots or robots that 
  are sent to space missions where efficiency matters. Furthermore, we have used the framework to estimate the 
  parameters of a desired system but it can also be used for adaptive nonlinear control applications --- where 
  the controller needs to adapt to unknown changes in dynamics and kinematics of robots \citep{cheah06,dewolf16} --- 
  to give an efficient adaptive spiking controller. The current framework is obtained for deterministic dynamical 
  systems -- future work will extend it to stochastic dynamical systems.

%\bibliography{nonlin_dyn}

%
\subsubsection*{Acknowledgments}

%
%\section*{References}
%
%References follow the acknowledgments. Use unnumbered first-level
%heading for the references. Any choice of citation style is acceptable
%as long as you are consistent. It is permissible to reduce the font
%size to \verb+small+ (9 point) when listing the references. {\bf
%  Remember that you can use a ninth page as long as it contains
%  \emph{only} cited references.}
%\medskip
%
%\small
%
%[1] Alexander, J.A.\ \& Mozer, M.C.\ (1995) Template-based algorithms
%for connectionist rule extraction. In G.\ Tesauro, D.S.\ Touretzky and
%T.K.\ Leen (eds.), {\it Advances in Neural Information Processing
%  Systems 7}, pp.\ 609--616. Cambridge, MA: MIT Press.
%
%[2] Bower, J.M.\ \& Beeman, D.\ (1995) {\it The Book of GENESIS:
%  Exploring Realistic Neural Models with the GEneral NEural SImulation
%  System.}  New York: TELOS/Springer--Verlag.
%
%[3] Hasselmo, M.E., Schnell, E.\ \& Barkai, E.\ (1995) Dynamics of
%learning and recall at excitatory recurrent synapses and cholinergic
%modulation in rat hippocampal region CA3. {\it Journal of
%  Neuroscience} {\bf 15}(7):5249-5262.

\end{document}